\UseRawInputEncoding 
\documentclass[10pt,journal,compsoc]{IEEEtran}

\usepackage{booktabs}
\usepackage{graphicx}
\usepackage[table,xcdraw,x11names,dvipsnames]{xcolor}
\usepackage{dblfloatfix}
\newcommand{\todo}[1]{}
\renewcommand{\todo}[1]{{\color{red} TODO: {#1}}}
\usepackage{comment}
\usepackage{amssymb}
\usepackage{multirow}
\usepackage{lscape}
\usepackage{array, multirow, bigdelim, makecell}
\usepackage{fontawesome5}

\usepackage[skins]{tcolorbox} 
\usepackage{cuted}
\usepackage{enumitem}

\newcommand*\circled[1]{\tikz[baseline=(char.base)]{
            \node[shape=circle,fill,inner sep=0.9pt] (char) {\textcolor{white}{#1}};}}

\def\mybarhhigh#1#2{
   {\color{black}\rule{#1mm}{8pt}}  #2}
   
\def\mybar2#1#2#3#4#5{
   {\color{black}\rule{4pt}{#1mm}}
 {\color{black}\rule{4pt}{#2mm}}
  {\color{black}\rule{4pt}{#3mm}}
   {\color{black}\rule{4pt}{#4mm}}
    {\color{black}\rule{4pt}{#5mm}}
   }

\def\mybarZ#1#2#3#4#5#6#7{
   {\color{black}\rule{4pt}{#1mm}}
 {\color{black}\rule{4pt}{#2mm}}
  {\color{black}\rule{4pt}{#3mm}}
   {\color{black}\rule{4pt}{#4mm}}
    {\color{black}\rule{4pt}{#5mm}}
     {\color{black}\rule{4pt}{#6mm}}
      {\color{black}\rule{4pt}{#7mm}}
   }
   
\ifCLASSINFOpdf

\else

\fi

\begin{document}
%
\title{ The Role of Emotional Intelligence in Handling Requirements Changes in Software Engineering}
%
%
%
%

\author{Kashumi~Madampe,~\IEEEmembership{Member,~IEEE,}
        Rashina~Hoda,~\IEEEmembership{Member,~IEEE,}
        and~John~Grundy,~\IEEEmembership{Senior Member,~IEEE}
\IEEEcompsocitemizethanks{\IEEEcompsocthanksitem K. Madampe, R. Hoda, and J. Grundy are with HumaniSE Lab, Department of Software Systems and Cybersecurity, Monash University, Australia. 
\protect\\
E-mail: kashumi.madampe@monash.edu
}
\thanks{Manuscript received June, 2022; revised xxxx.}}

%
%

\markboth{Submitted to IEEE Transactions on Software Engineering}%
{Madampe \MakeLowercase{\textit{et al.}}: Bare Advanced Demo of IEEEtran.cls for IEEE Computer Society Journals}
%



\IEEEtitleabstractindextext{%
\begin{abstract}
\textbf{Background: }Requirements changes (RCs) are inevitable in Software Engineering. Research shows that \textit{emotional intelligence} (EI) should be used alongside agility and cognitive intelligence during RC handling.
\textbf{Objective: }We wanted to study the role of EI in--depth during RC handling.
\textbf{Method: }
We conducted a socio--technical grounded theory study with eighteen software practitioners from Australia, New Zealand, Singapore, and Sri Lanka.
\textbf{Findings: }
We found causal condition (software practitioners handling RCs), intervening condition (mode of work), causes (being aware of own emotions, being aware of others' emotions), direct consequences (regulating own emotions, managing relationships), extended consequences (sustaining productivity, setting and sustaining team goals), and contingencies: strategies (open and regular communication, tracking commitments and issues, and ten other strategies) of using EI during RC handling. We also found the covariances where strategies co-vary with the causes and direct consequences, and ease/ difficulty in executing strategies co-vary with the intervening condition. 
\textbf{Conclusion: }Open and regular communication is key to 
EI during RC handling. To the best of our knowledge, the framework we present in this paper is the first theoretical framework on EI in Software Engineering research. We provide recommendations including a problem--solution chart in the form of causes, direct consequences, and mode of work against the contingencies: strategies 
for software practitioners to consider during RC handling, and future directions of research.
\end{abstract}

}

\maketitle

\IEEEdisplaynontitleabstractindextext

%
\IEEEpeerreviewmaketitle

\ifCLASSOPTIONcompsoc
\IEEEraisesectionheading{\section{Introduction}\label{sec:introduction}}
\else
\section{Introduction}
\label{sec:introduction}
\fi

\IEEEPARstart{A}{s} agile methods become the predominant approach to Software Engineering, the frequent introduction of requirements changes (RCs) -- such as additions, modifications, and deletions of functional and non-functional requirements in the software development process -- has also become common \cite{Madampe2021AContexts}.
From receiving, developing and testing to delivering of the RCs, software practitioners use various techniques \cite{Madampe2020TowardsTeamsb}, and show several emotions, such as \textit{excitement, anxiety,} and \textit{fatigue}, during the RC handling life cycle \cite{Madampe2022TheEngineering}.

Emotional intelligence (EI) is the ability to process emotional information and use it in reasoning and other cognitive activities \cite{VandenBos2007APAPsychology}. Our previous studies show that for effective handling of these RCs it is not sufficient to rely only on agility and cognitive intelligence (one's abilities to learn, remember, reason, solve problems, and make sound judgements \cite{VandenBos2007APAPsychology}), but EI is needed as well.
To the best of our knowledge there exists no literature on EI in handling RCs. Given its vital importance, we decided to study this in depth, hence the central focus of this study is:
\begin{center}
    \textbf{``The role of software practitioners' emotional intelligence during requirements changes handling''}
\end{center}

For our investigation, we used a Socio--technical Grounded Theory (STGT) for data analysis  \cite{Hoda2021Socio-TechnicalEngineeringb}. STGT is an ideal approach to study complex undiscovered phenomena in socio--technical contexts such as ours. The data collection and analysis were iterative (2 iterations: \textit{iteration 1} (\circled{1}), \textit{iteration 2} (\circled{2})) and interleaved. We collected qualitative data by conducting eighteen semi-structured face-to-face and online interviews (\circled{1}\footnote{Approved by The University of Auckland Human Participants Ethics Committee. Approval Number: 023015}: ten interviews, \circled{2}\footnote{Approved by Monash Human Research Ethics Committee. Approval Number: 23578}: eight interviews) with software practitioners from Australia (N=8), New Zealand (N=8), Sri Lanka (N=1), and Singapore (N=1). The interviews in \circled{1} lasted approximately 60 minutes, and the interviews in \circled{2} lasted approximately 30 minutes. The interviews were also accompanied by pre--interview questionnaires (\circled{1}: 10 minutes to fill; \circled{2}: 30 minutes to fill) to gather data on the context of the participants. 

We analysed data using STGT data analysis techniques such as open coding, constant comparison, and memoing. We identified that EI plays an important role in RC handling. Based on our analysis, we describe the six C's associated with our study focus: the context, condition, causes, consequences, contingencies (or strategies), and co-variances related to EI in RC handling that we present as a theoretical framework to guide research and practice. The key contributions of our work are:

\begin{itemize}
    \item Adding new knowledge to the growing literature on human aspects in software engineering by explaining the phenomenon of software practitioners' \textit{emotional intelligence during RC handling};
    \item Documenting the most common strategies whereby software practitioners use their emotional intelligence during RC handling;
    \item Presenting a set of practical recommendations derived from our findings for SE practitioners; and
    \item Identifying some key future research directions for SE researchers.
\end{itemize}

\section{Definitions, Background, and a Motivating Example}

\subsection{Definitions}
We use some specialist terminology throughout our paper. We define them as below (cited definitions are from direct sources, and uncited definitions are our own). The terms are listed in their order of appearance in the paper.
\begin{itemize}[leftmargin=*]
\item \textbf{Emotion:} A sequence of interrelated, synchronised changes in the states of all the five organismic subsystems (information processing, support, executive, action, and monitoring) in response to the evaluation of an external or internal stimulus event as relevant to central concerns of the organism \cite{Scherer1987TowardEmotion}
\item \textbf{Emotional intelligence: }Type of intelligence that involves the ability to  process emotional information and use it in reasoning and other cognitive activities  \cite{VandenBos2007APAPsychology}
\item \textbf{Emotion regulation: }Any process that decreases, maintains, or increases emotional intensity over time, thereby modifying the spontaneous flow of emotions \cite{Koval2015EmotionInertia}, \cite{Gross2007EmotionPress.}, \cite{Koole2009TheReview}
\item \textbf{Emotional response: }An emotional reaction, such as happiness, fear, or sadness, to give a stimulus  \cite{VandenBos2007APAPsychology}  
\item \textbf{Empathy: }Understanding a person from his or her frame of reference rather than one's own, or vicariously experiencing that person's feelings, perceptions, and thoughts \cite{VandenBos2007APAPsychology}
\item \textbf{Central phenomenon:} The focus of the study \cite{Glaser1967}
    \item \textbf{Context:} This is where the central phenomenon took place. This includes participants' demographics, and their team and project information \cite{Glaser1967}
    \item \textbf{Condition:} A factor that affects the central phenomenon \cite{Glaser1967}
    \item \textbf{Causal condition:} The condition that gave rise to the central phenomenon \cite{Glaser1967}
    \item \textbf{Intervening condition:} A condition that alters the central phenomenon in some way \cite{Strauss1998BasicsTechniques.}
    \item \textbf{Cause:} The result of the central phenomenon \cite{Glaser1967}
    \item \textbf{Consequence:} The output of the cause \cite{Glaser1967}
    \item \textbf{Direct consequence:} The immediate output of the cause
   \item \textbf{Extended consequence:}The immediate output of the direct consequence
    \item \textbf{Contingency:} Any event that can occur given a specific cause or condition \cite{Glaser1967}
    \item \textbf{Strategy:} An action that can be executed
    \item \textbf{Variable:} Context, condition, cause, consequence, or contingency which are the elements of the central phenomenon \cite{Glaser1967}
    \item \textbf{Covariance:} The change of one variable with another \cite{Glaser1967}
\end{itemize}

\subsection{Background}

\subsubsection{Requirements Change Handling}
Agile methods are widely used in software engineering contexts and promote the idea of introducing RCs even late in development \cite{Beck2001ManifestoDevelopment}. However, software practitioners encounter many challenges such as lack of requirements traceability \cite{Inayat2015}, incorrect requirements prioritisation \cite{Inayat2015}, minimal requirements documentation \cite{Madampe2021AContexts}, \cite{Bjarnason2011AEngineeringb}, \cite{Inayat2015}, \cite{Kapyaho2015AgileStudy}, contractual issues \cite{Inayat2015}, and customer agreement \cite{Inayat2015} in handling these new RCs. 

In addition to these challenges, in our earlier work  we found that RCs in agile contexts are challenging to handle when their complexity is high, cascading impact is high, size is large, the effort required is high, the definition is imprecise or unclear, the priority is high, the access to customer is difficult or irregular, and the cross-functionality is forced \cite{Madampe2022Emotion-CentricEngineering}. Even though there are practices (e.g.,  face--to--face communication \cite{Bakalova2011AgileLiterature}, \cite{Baruah2015}, \cite{DeOliveiraNeto2017ChallengesStudy}, \cite{Jun2010ApplicationDevelopment} iterative requirements \cite{Albuquerque2020DefiningStudy}, \cite{Ramesh2010AgileStudyc}, \cite{Jun2010ApplicationDevelopment} \cite{DeOliveiraNeto2017ChallengesStudy}, prototyping \cite{Albuquerque2020DefiningStudy}, \cite{Ramesh2010AgileStudyc}, \cite{Wagner2018AgileProblems} review meetings \cite{Ramesh2010AgileStudyc}, \cite{DeOliveiraNeto2017ChallengesStudy}, and prioritisation \cite{Albuquerque2020DefiningStudy}, \cite{Ramesh2010AgileStudyc}, \cite{DeOliveiraNeto2017ChallengesStudy}, see Table~\ref{tab:project_info}) that are available to mitigate these challenges and handle RCs, most of these challenges result in varying emotional responses in software practitioners \cite{Madampe2022TheEngineering}, \cite{Madampe2022Emotion-CentricEngineering}.

\subsubsection{Emotional Intelligence}
 There are several definitions for EI. Salovey and Mayer, who proposed the concept of EI, define it as ``a type of intelligence that involves the ability to process emotional information and use it in reasoning and other cognitive activities'' \cite{VandenBos2007APAPsychology}. This concept was made popular by Daniel Goleman, who proposed in his work with his colleagues \cite{Goleman2002TheTeams} the key aspects of EI, such as \textit{self-awareness} -- awareness of own emotions, \textit{social awareness} -- awareness of others' emotions, \textit{self-management} -- regulating own emotions, and \textit{relationship management} -- managing relationships. The findings on causes and direct consequences we present in this paper were named using these four aspects. As researchers found assessing EI is worthy of attention, several EI assessment measures currently exist. These EI measures fall into ability--based or mixed--based model categories. The EI measures that are currently available and their details are summarised online\footnote{https://www.eiconsortium.org/measures/measures.html}, hence we do not explain about them here. 
 
In our earlier work we found that agility, cognitive intelligence, and EI of SE practitioners need to be used together to handle RCs effectively \cite{Madampe2022Emotion-CentricEngineering}. Studies focusing on agility and cognitive intelligence are prominent in Software Engineering research. However, studies focusing on EI are very limited (\cite{Rezvani2019EmotionalProjects}, \cite{Kosti}). Kosti et al. \cite{Kosti} show that there are connections between the EI of the developers and their work preferences. They say that, people with higher EI prefer being responsible of the entire development process, and like to prioritise the tasks by themselves rather than a manager doing that for them. According to \cite{Rezvani2019EmotionalProjects}, the influence of EI on software developers' stress is negative, and EI fosters trust among developers. As yet, there exist no studies focusing exclusively on EI during RC handling.

\subsection{Motivating Example}

Consider Kash, who is a developer in a software team. She receives a new RC to work with her colleagues. Because the change was unexpected and impacts her current work, she feels \textit{frustrated} about this new work (awareness of own emotions). She also perceives that for similar reasons her teammates are also \textit{frustrated} (awareness of others' emotions). She then thinks it is better to \emph{share how she feels} with her team during their retrospective meeting (a strategy for being aware of own emotions, and regulating own emotions). She understands that she has to be \emph{open to new RCs} as she is working in a software development context (a strategy to regulate her own emotions). She also thinks that if herself and her team have \emph{open and regular communication} about new RCs then it will help everyone to bond and have better relationships within the team (a strategy to manage team relationships). Furthermore, she finds that  these approaches allow her to \emph{sustain her own productivity} within the team, but also to \emph{set and sustain team goals}.
However, when working in an office environment (in--person), she finds that some strategies that she uses to be aware of her emotions and her teammates' emotions, to regulate her emotions, and to manage relationships with teammate, are easier to apply in--person than remotely. 

\begin{figure*}
    \centering
    \includegraphics[scale=0.35]{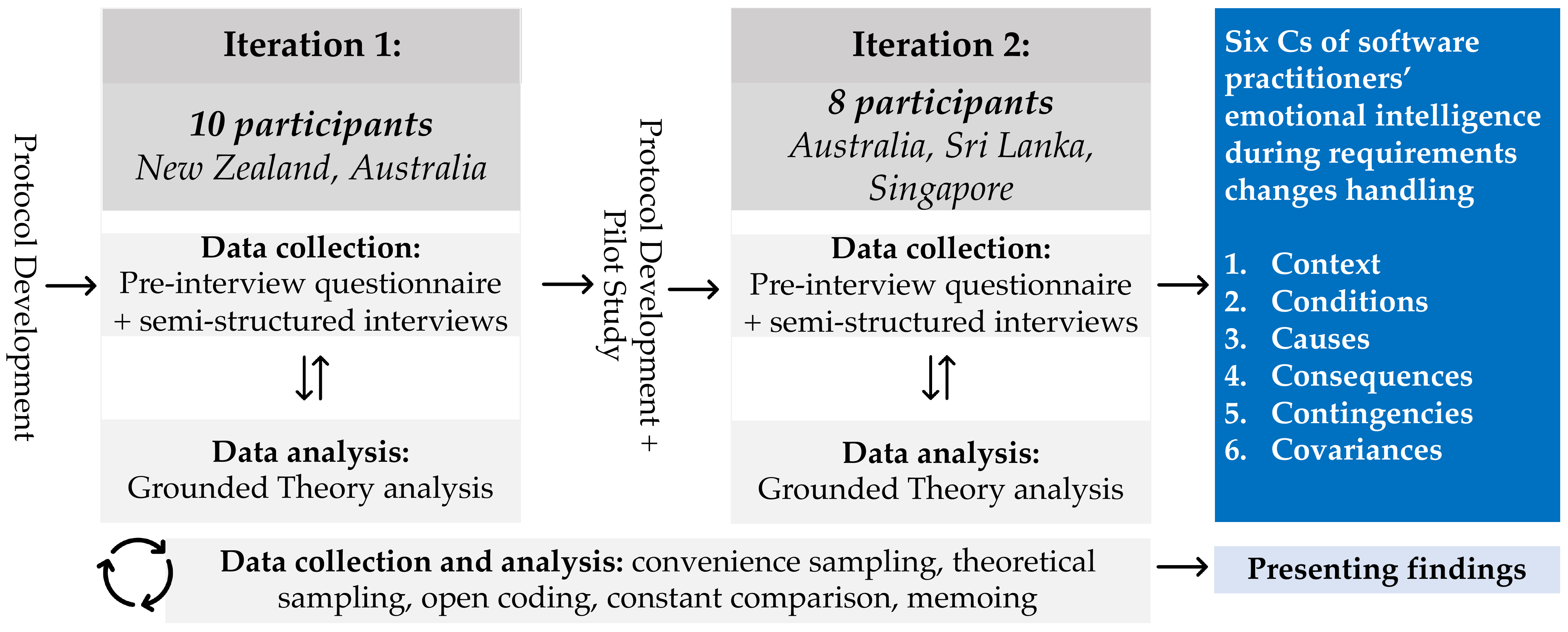}
    \caption{STGT Study Approach (Protocol development includes pre--interview questionnaire development, interview guide formation, and explanatory statement and advertisement preparation)}
    \label{fig:approach}
\end{figure*}

\section{Approach}

An overview of our study approach is given in Fig. \ref{fig:approach}. We collected data from practitioners in Australia, New Zealand, Singapore, and Sri Lanka in two iterations (one for high level understanding of general issues in RC handling and emotional responses to RCs, and the other with a finer focus on emotional intelligence). We used socio--technical grounded theory (STGT) \cite{Hoda2021Socio-TechnicalEngineeringb} techniques such as open coding, constant comparison, and memoing to analyse the data. The steps we followed consisted of developing the study protocol (pre--interview questionnaire development, interview guide formation), conducting a pilot study, collecting data, and analysing data. We elaborate on these steps in the following sub--sections. 

	\subsection{Protocol Development}
	
    \textbf{Pre--interview Questionnaire Development.}
			We collected participants' demographics (Section \ref{sec:demo}), team, and project information (Section \ref{sec:project}), through a pre--interview questionnaire. Having a pre--interview questionnaire helped us gain a better understanding of the participants and the context they are planning to discuss in the interviews in advance, and also allowed us to have enough time to focus the interview on the RC handling situation. The questions in the pre--interview questionnaires of \circled{1} and \circled{2} were the same, except that \circled{2} had an additional question about the impact of Covid--19 on RC handling, and an EI scale. 
			All the pre--interview questions were mandatory and close--ended except for the question on Covid--19, which was an optional open--ended question. The approximate time spent on filling the pre--interview questionnaire per participant was 30 minutes in \circled{2},
			and 10 minutes in \circled{1}. As we recruited participants from a software company in \circled{2}, we had an additional step where a senior management level employee reviewed the pre--interview questionnaire to ensure the questions were in line with company policies.
			
			\textbf{Interview Guide Design.}
			To get an in--depth understanding of the role of EI in handling RCs, we collected qualitative data through semi--structured interviews. One advantage of conducting interviews was the ability to ask the participants follow up questions, which is a limitation in open--ended questions in surveys. To guide our interviews, we designed the interview guides (one for each iteration) with an estimated time of 60 minutes and 30 minutes respectively. The interview guide in \circled{1} focused on general issues of emotions experienced during RC handling, and \circled{2} focused solely on EI during RC handling. 
			
			In \circled{1}, we started the interview by asking the participants to walk through a scenario where they had RCs, and then we asked follow-up questions such as ``how did you feel when you received the RC?'' and so on. As the study progressively focused on EI, in \circled{2}, we included questions to cover the four aspects of EI as defined by Goleman et al. \cite{Goleman2002TheTeams}, i.e., self-awareness, self-management, social awareness, and relationship management. This was done to ensure the proper treatment of EI as defined in Psychology. However, the actual terms were not used in the interview questions and interviewees were allowed to express their experiences freely. Therefore, in the interview guide of \circled{2}, we first asked the participant to describe the situation that they used when answering the pre--interview questionnaire. Then we designed the rest of the questions to elicit more data to answer our research questions. For example, the starting question ``Let's talk about the RC handling situation you used to answer the questions in the pre--interview questionnaire. Can you describe it for me?'' followed by questions such as ``how did that make you feel?,'' ``what did you do to manage how you felt?,'' ``how did your teammates react?,'' ``how did you work as a team in that situation?,'' “what could have been different for you not to experience so?'' (if the situation was challenging) and ``how did the feelings you had affect the progress of working on that RC?.'' We also prepared a slide with a list of emotions, similar to \cite{Madampe2022TheEngineering}, to help participants recall how they felt in answering questions related to how they felt when handling RCs.

\textbf{Explanatory Statement and Advertisement Preparation.}
\textit{Explanatory statement. }In both iterations, we documented the details of the study in a statement in which we explained what the  participants are expected to do, and further crucial information such as the impact of participation/non--participation/withdrawal of participation in the study, and confidentiality.

\textit{Advertisement.} We designed an advertisement to gauge the interest of potential participants. The advertisement provided a high level view of the study. Especially since \circled{2} focused on EI, both the advertisement and the explanatory statement avoided the words ``emotions'' and ``intelligence'' to reduce the drop--rate of potential participants and any biases that could occur when answering the questions.

		\subsection{Transitional Piloting} %
		As we transitioned from \circled{1}, asking more generally about emotions when handling RCs, to \circled{2} with a deeper focus on EI, we decided to pilot our revised protocols with an industry practitioner. An experienced software architect from Sri Lanka helped us pilot our revised protocols. It became apparent that if the pre--interview questionnaire was filled in advance, the participant was likely to forget it by the interview time. Due to this experience, we decided to ask the participants in our study to fill in the pre--interview questionnaire an hour before the scheduled interview time. This made us change our initial scheduling plan in \circled{2} by having two blocks scheduled for the participants -- one to fill in the pre--interview questionnaire, and the other for the interview.

		\subsection{Data Collection}
		The study commenced with the use of convenience sampling and later moved to using theoretical sampling.
		
			\textbf{Participant Recruitment.}
			In \circled{1}, we made an open call by sharing the advertisement with the explanatory statement on social media such as LinkedIn, Twitter and Facebook. We recruited ten participants in this iteration. In \circled{2}, we sent the same artefacts (but with different content) to personal contacts and also recruited six participants from a large software company. An employee who works at this company shared the advertisement and the explanatory statement with the potential participants. They also helped in scheduling the pre--interview questionnaire filling time and interviews. 
			The participant demographics, project and team information, and the scenarios participants used in the study are given in Section \ref{sec:findings}.

			\textbf{Distribution of the Pre--interview Questionnaire.}
			The first author shared a link to the pre--interview questionnaire with potential participants who showed their interest in participating in the study one or two days before the interview. In \circled{1}, the pre--interview was hosted by Google Forms at The University of Auckland. The pre--interview questionnaire of \circled{2} was hosted on Qualtrics at Monash University. The first author made sure that pre--interview questionnaires were not filled way ahead of the interview, but only just before the interview, by checking Google Forms and Qualtrics.

			\textbf{Conducting Semi--structured Interviews.}
			In \circled{1}, the first author conducted the seven interviews face--to--face, and three interviews online (Skype). In \circled{2}, the first author conducted all the interviews online (Zoom) with the participants due to the Covid--19 pandemic. The average time spent on interviews remained the same as we planned (60 minutes and 30 minutes respectively). 

		\subsection{Data Analysis}
			In \circled{1}, the audio-recorded interviews were transcribed by the first author and a professional transcriber under confidential agreement. In \circled{2}, we used otter.ai\footnote{https://otter.ai/} to transcribe the interviews. Furthermore, MAXQDA was used to analyse the qualitative data. We analysed the qualitative data collected from the interviews and from the open--ended question on Covid--19 using open coding and constant comparison techniques in Socio--Technical Grounded Theory (STGT) \cite{Hoda2021Socio-TechnicalEngineeringb}. We used STGT data analysis techniques as STGT is designed to uncover the insights in socio--technical environments, and due to positive experience using it in our previous studies \cite{Madampe2022TheEngineering}, \cite{Madampe2022Emotion-CentricEngineering}. 
			
\begin{figure*}
			    \centering
			    \includegraphics[scale=0.4]{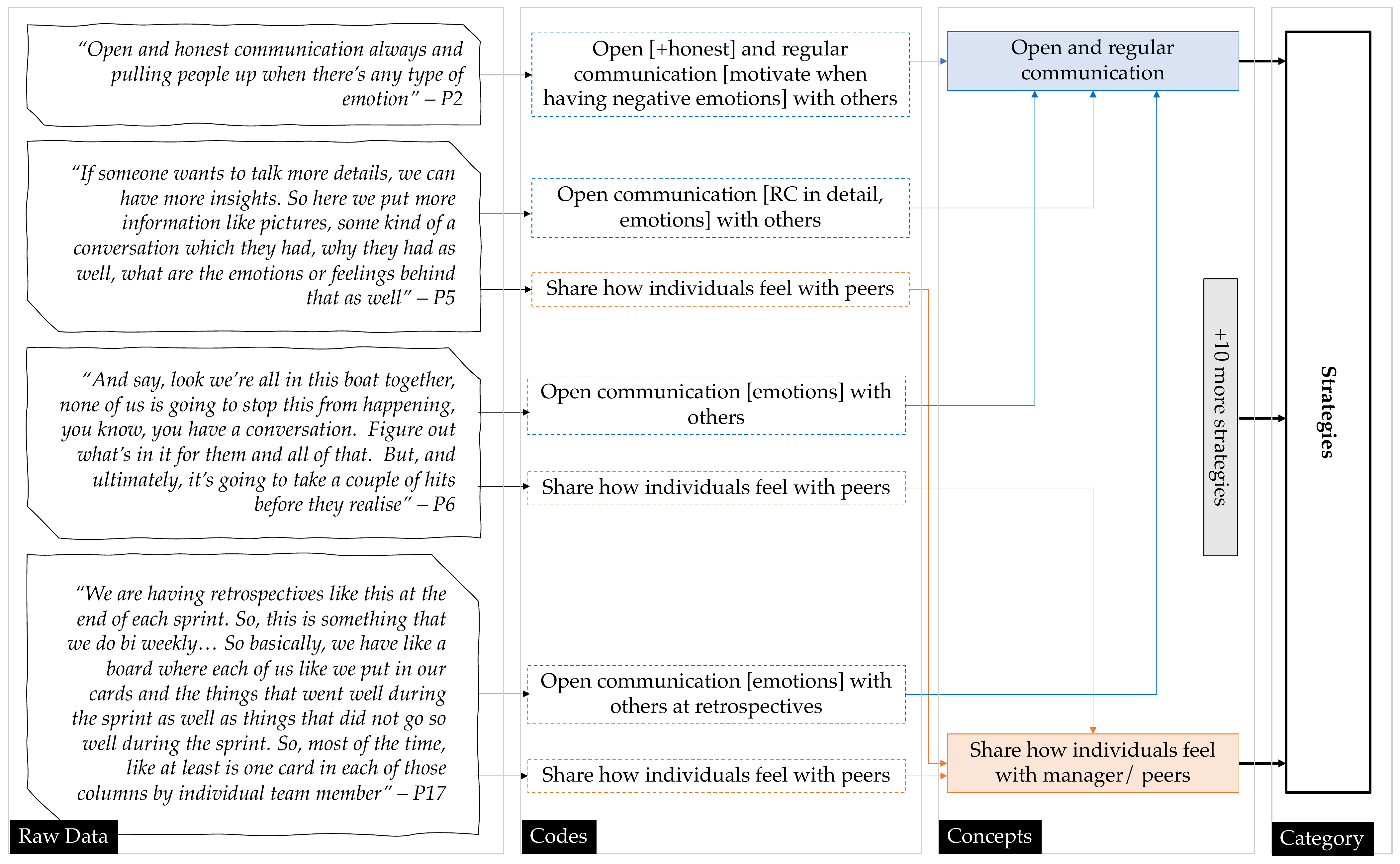}
			    \caption{Emergence of the Category: Strategies from raw data $\rightarrow$ codes $\rightarrow$ concepts $\rightarrow$ category through constant comparison}
			    \label{fig:data_analysis}
			\end{figure*}

			\textbf{Open Coding and Constant Comparison. }Fig. \ref{fig:data_analysis} shows an example of STGT analysis. In this example, the raw data ``\textit{Open and honest communication always and pulling people up when there's any type of emotion}'' resulted in the code ``Open [+honest] and regular communication [motivate when having negative emotions] with others,'' which is the researcher's interpretation of data in small chunks of meaningful words. These codes were then constantly compared, and similar codes were grouped to develop the concept ``Open and regular communication.'' The same constant comparison technique was applied to the concepts to come up with the category ``Strategies.'' 
			
			\textbf{Memoing.} Memos were written in detail as interview reflections (See Appendix for an example), especially in \circled{2}. Conceptual memos (See Appendix for an example) were written throughout the data collection and analysis process. Memoing helped in identifying relationships between categories. All findings were then visualised in Miro (visual memo). 
	        
	        \textbf{Final Output -- A Theoretical Framework.} Together, the above led to the development of a theoretical framework presented in Fig. \ref{fig:sixCs}. To visualise our framework, we consulted Glaser's six Cs model \cite{Glaser1967} and Strauss and Corbin's paradigm model \cite{Strauss1998BasicsTechniques.}.

\section{Findings}
\label{sec:findings}
\begin{figure*}
    \centering
    \includegraphics[width=\textwidth]{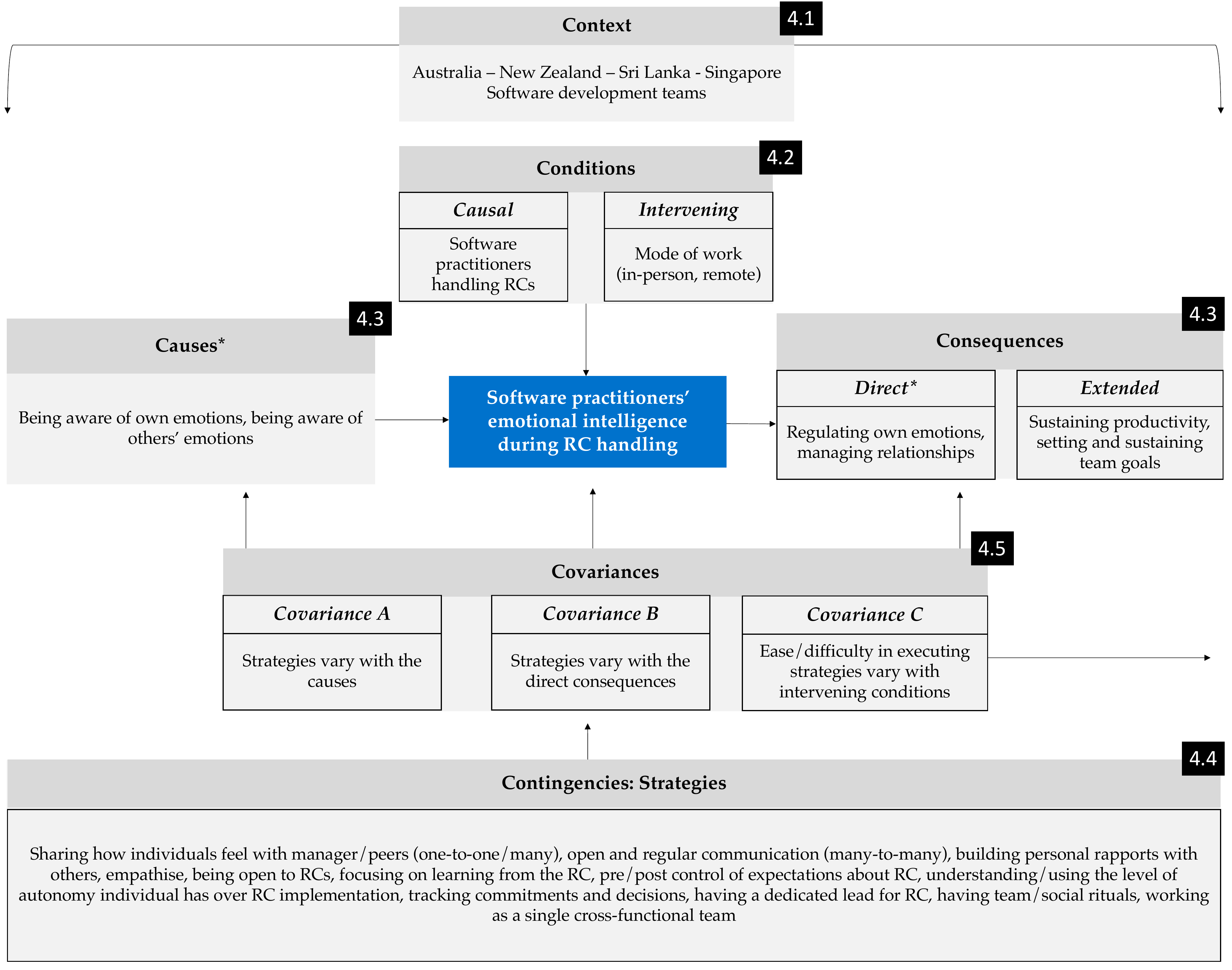}
    \caption{Six Cs of Software Practitioners' Emotional Intelligence during RC Handling (*These are four aspects of Emotional Intelligence as defined by Goleman et al. \cite{Goleman2002TheTeams}, which we used to name the concepts. They are depicted on different sides of the central phenomenon because of the relationships among them.)}
    \label{fig:sixCs}
\end{figure*}

Within the context of software teams in Australia, New Zealand, Sri Lanka, and Singapore, we found conditions, causes, consequences, contingencies: strategies, and covariances, which altogether are the six Cs of software practitioners' Emotional Intelligence during Requirements Change handling. The six Cs are shown in Fig. \ref{fig:sixCs}. In this section, first we explain the context, then the conditions, and after that causes and consequences, contingencies, and covariances in order.

\subsection{Context}
\subsubsection{Participant Demographics} 
\label{sec:demo}
Eighteen software practitioners participated in our study. We had an equal number of participants from Australia and New Zealand (N=8; 44.44\% each). The remaining two participants were from Sri Lanka and Singapore. 10\% (N=6) of our participants played more than one role in their projects. They included managers and Scrum Masters (N=5; 27.78\% each); developers (N=4; 22.22\%); and other roles such as Business Analyst, Product Owner, Tester, Senior Solutions Architect/ Principle Architect, and Global Head of Projects. They had a mean total experience of 14.39 years (min--total--experience=2 years; max--total--experience=56 years), and a mean agile experience of 6.81 years (min--agile--experience=1; max--agile--experience=18 years). The detailed demographic information of the participants is given in Table \ref{tab:demographics}.

\begin{table*}[t]
\caption{Demographic Information of the Participants (P\#: Participant ID; XT: Total Experience (years); XTA: Total agile experience (years); NZ: New Zealand; AU: Australia; SG: Singapore; SL: Sri Lanka; XP: Extreme Programming; FDD: Feature Driven Development; DSDM: Dynamic Systems Development Method; TDD: Test Driven Development.)}
\label{tab:demographics}
\resizebox{\textwidth}{!}{%
\begin{tabular}{@{}llllll@{}}
\toprule
\textbf{P\#} & \textbf{Country} & \textbf{XT}   & \textbf{XTA} & \textbf{Agile Methods Experience}                                                          & \textbf{Role in the Project}                             \\ \midrule
P1  & NZ      & 20   & 12  & Scrum, XP, Scrum XP combo, Kanban, FDD, Scrumban            & Tester/ Scrum Master                            \\
P2  & NZ      & 56   & 16  & Scrum, XP, Scrum XP combo, Kanban                           & Manager                                         \\
P3  & NZ      & 24   & 4   & Scrum, Kanban                                               & Business Analyst                                \\
P4  & AU      & 25   & 8   & Scrum, XP, Kanban, FDD, DSDM                                & Global Head of Projects                         \\
P5  & NZ      & 15   & 4.6 & Scrum, XP, Scrum XP combo                                   & Scrum Master                                    \\
P6  & NZ      & 26   & 1   & Scrum                                                       & Scrum Master                                    \\
P7  & NZ      & 3    & 2   & Scrum, Kanban                                               & Scrum Master                                    \\
P8  & NZ      & 19.5 & 18  & Scrum, XP, Scrum XP combo, Kanban, Spotify, Company methods & Senior Solutions Architect/ Principle Architect \\
P9  & NZ      & 7    & 5   & Scrum, Kanban, FDD                                          & Tester                                          \\
P10 & AU      & 2    & 2   & Scrum, Kanban                                               & Scrum Master/ Business Analyst/ Manager         \\
P11 & SG      & 6    & 4   & Scrum                                                       & Developer/ Product Owner                        \\
P12 & SL      & 3    & 3   & Scrum, Scrum and XP combo, Kanban                           & Agile Coach/ Scrum Master                       \\

P13 & AU      & 8    & 8   & Scrum, Scrum XP comobo, Kanban, FDD, Spotify, Waterfall     & Manager                                         \\
 
P14 & AU      & 11   & 11  & Scrum, Kanban, Waterfall                                    & Developer                                       \\

P15 & AU      & 6.5  & 6.5 & Scrum                                                       & Manager                                         \\

P16 & AU      & 19.5 & 10  & Scrum, Kanban, FDD, Waterfall                               & Manager/ Developer/ Tester                      \\

P17 & AU      & 3.5  & 3.5 & Scrum, TDD                                                  & Developer                                       \\

P18 & AU      & 4    & 4   & Scrum, Kanban                                               & Developer                                       \\ \bottomrule
\end{tabular}%
}
\end{table*}

\subsubsection{Participants' Team and Project Information} 
\label{sec:project}
While all of our participants were working in team contexts, they used examples from their current/most recent projects to share their experiences with us. All participants used agile methods in their projects. The team and project information are summarised in Table \ref{tab:project_info}.

\begin{table*}[]
\caption{Information of Current/ Most Recent Project of the Participants (Other: Product Support and Rewrite/Update of Existing Functionality; XP: Extreme Programming; DSDM: Dynamic Software Development Method; Some Participants' Projects Belonged to More Than One Category, and Some Participants Used More Than One Software Development Method)}
\label{tab:project_info}
\begin{tabular}{@{}llllll@{}}
\toprule
\textbf{Project Domain}     & {\textbf{\# of Participants}} & \textbf{Project Category} & \textbf{\# of Participants} & \textbf{Development Method Used} & {\textbf{\# of Participants}} \\ \midrule
Transport                   & \mybarhhigh{1}{1}                                               & Software as a Service     & \mybarhhigh{3}{3}       & XP                               & \mybarhhigh{4}{4}                                               \\
IT                          & \mybarhhigh{11}{11}                                              & New Development           & \mybarhhigh{10}{10}      & Kanban                           & \mybarhhigh{6}{6}                                               \\
Finance \& Banking          & \mybarhhigh{3}{3}                                               & Migration                 & \mybarhhigh{4}{4}      & Scrum                            & \mybarhhigh{11}{11}                                              \\
Business Services           & \mybarhhigh{1}{1}                                               & Maintenance               & \mybarhhigh{6}{6}       & DSDM                             & \mybarhhigh{1}{1}                                               \\
Facilities                  & \mybarhhigh{1}{1}                                               & Other                     & \mybarhhigh{2}{2}       & ScrumKanban Combo                & \mybarhhigh{1}{1}                                               \\
Media and Communications    & \mybarhhigh{1}{1}                                               &                           &                             & ScrumXP Combo                    & \mybarhhigh{1}{1}                                               \\
                            & \multicolumn{1}{l}{}                            &                           &                             & ScrumWaterfall Combo             & \mybarhhigh{1}{1}                                               \\ \cmidrule(r){1-2} \cmidrule(l){5-6} 
\textbf{Team Size}          & \multicolumn{1}{l}{\textbf{\# of People}}       &                           &                             & \textbf{Iteration Length}        & \multicolumn{1}{l}{\textbf{\# of Weeks}}        \\ \cmidrule(r){1-2} \cmidrule(l){5-6} 
Minimum                     & 4                                               &                           &                             & Minimum                          & 1                                               \\
Maximum                     & 50                                              &                           &                             & Maximum                          & 5                                               \\
Mean                        & 10.61                                           &                           &                             & Mean                             & 2                                               \\
Standard Deviation          & 10.56                                           &                           &                             & Standard Deviation               & 0.84                                            \\ \midrule
\multicolumn{6}{l}{\textbf{Practices Followed (Order of the bars in each graph below: Never $\rightarrow$ Sometimes $\rightarrow$ About half of the Time $\rightarrow$ Most of the Time $\rightarrow$ Always)}}                                                                      \\ \midrule
Collective Estimation       & \mybar2{0}{0.4}{0.2}{1.2}{1.2}                & Product Backlog           & \mybar2{0}{0.2}{0.6}{0.4}{2.4}                  & Scrum/ Kanban Board              & \mybar2{0}{0.2}{0}{0.4}{3}               \\
Customer Demos              & \mybar2{0.6}{0.6}{0.4}{1}{1}                 & Short Iterations/ Sprints & \mybar2{0}{0}{0.2}{0.8}{2.6}                  & Self-assignment                  & \mybar2{0.2}{0.2}{0.6}{0.6}{2}                 \\
Daily Standup/ Team Meeting & \mybar2{0}{0}{0.2}{0.4}{3}                 & Release Planning          & \mybar2{0.2}{0.8}{0.4}{1.4}{0.8}                  & Sprint Backlog                   & \mybar2{0.2}{0.4}{0.4}{0.2}{2.4}                \\
Definition of Done          & \mybar2{0.4}{0}{0.4}{1}{1.8}               & Retrospectives            & \mybar2{0}{0}{0.4}{0.8}{2.4}                 & User Stories                     & \mybar2{0.2}{0.6}{0.4}{0.8}{1.6}                 \\
Iteration Planning          & \mybar2{0}{0.2}{0.4}{1}{2}                  & Review Meetings           & \mybar2{0.4}{0.2}{0.4}{0.8}{1.8}         & Pair Programming                 & \mybar2{0.6}{1.4}{0.2}{1}{0.4}               \\ \bottomrule
\end{tabular}
\end{table*}

\subsection{Conditions}

Our findings indicate that the EI of practitioners during RC handling 
comes to play under the \textit{causal} and \textit{intervening} conditions. The \emph{causal condition} we found was software practitioners handling RCs, and the \emph{intervening condition} we found was the mode of work.

\subsubsection{Causal Condition: Practitioners Handling RCs} 
\textit{--Shared by all participants.}

When filling the pre--interview questionnaire, each participant used a specific RC handling scenario from their current or most recent software development project. These scenarios caused several emotional responses in them, and emotions of others' around them, which they became aware of, regulated the emotions, and managed the relationships. Hence, practitioners' handling of RCs is the causal condition of our study. The majority of their scenarios were about modification of requirements (N=9), and the rest were additions (N=6), deletions (N=2), and a combination of a deletion and an addition (N=1). The summary of scenarios used by the participants and the causal condition derived from the scenarios are given in Table \ref{tab:conditions}.

\begin{table*}[]
\caption{Scenarios used in the interviews, respective causal, and intervening conditions}
\label{tab:conditions}
\resizebox{\textwidth}{!}{%
\begin{tabular}{@{}llll@{}}
\toprule
\textbf{P\#} & \textbf{Scenario used in the study}                                                                                                                                                                        & \textbf{\begin{tabular}[c]{@{}l@{}}Causal condition \\ (RC handling)\end{tabular}} & \textbf{\begin{tabular}[c]{@{}l@{}}Intervening condition \\ (mode of work)\end{tabular}} \\ \midrule
\rowcolor[HTML]{EFEFEF} 
P1           & \begin{tabular}[c]{@{}l@{}}Modifying a function after a customer requesting something different from \\ what was being started but was still aligning with sprint goal\end{tabular}                                                   & Modification                                                                       & In--person                                                                                \\
P2           & Modifying a function                                                                                                                                                                                           & Modification                                                                       & In--person                                                                                \\
\rowcolor[HTML]{EFEFEF} 
P3           & Modifying a function                                                                                                                                                                                & Modification                                                                       & In--person                                                                                \\
P4           & Adding new features to increase the business value after a market analysis                                                                                                                                     & Addition                                                                           & In--person                                                                                \\
\rowcolor[HTML]{EFEFEF} 
P5           & Modifying a function                                                                                                                                                                                           & Modification                                                                       & In--person                                                                                \\
P6           & Modifying a function                                                                                                                                                                                           & Modification                                                                       & In--person                                                                                \\
\rowcolor[HTML]{EFEFEF} 
P7           & Late addition of a missed requirement                                                                                                                                                                          & Addition                                                                           & In--person                                                                                \\
P8           & Modifying a feature after a release                                                                                                                                                                            & Modification                                                                       & In--person                                                                                \\
\rowcolor[HTML]{EFEFEF} 
P9           & \begin{tabular}[c]{@{}l@{}}Modifying a function after a customer requesting something different from \\ what was being started in the middle of a sprint\end{tabular}                                          & Modification                                                                       & In--person                                                                                \\
P10          & Adding a new feature after requesting a feature request                                                                                                                                                        & Addition                                                                           & In--person                                                                                \\
\rowcolor[HTML]{EFEFEF} 
P11          & Removal of features that end users no longer require                                                                                                                                                           & Deletion                                                                           & Remote                                                                                   \\
P12          & Adding multiple new requirements to upgrade an existing system                                                                                                                                                 & Addition                                                                           & In--person                                                                                \\
\rowcolor[HTML]{EFEFEF} 
P13          & Discovering new edge cases during development                                                                                                                                                                  & Addition                                                                           & Remote                                                                                   \\
P14          & \begin{tabular}[c]{@{}l@{}}Working on a modification due to a requirement that incepted considering \\ only the ideal product/ design case but not considering \\ the realities of engineering\end{tabular}    & Modification                                                                       & Remote                                                                                   \\
\rowcolor[HTML]{EFEFEF} 
P15          & \begin{tabular}[c]{@{}l@{}}Removal of a no longer necessary requirement that caused engineering \\ dilemmas\end{tabular}                                                                                       & Deletion                                                                           & Remote                                                                                   \\
P16          & \begin{tabular}[c]{@{}l@{}}Working on a modification due to the development of a requirement based \\ on wrong assumptions\end{tabular}                                                                        & Modification                                                                       & Remote                                                                                   \\
\rowcolor[HTML]{EFEFEF} 
P17          & \begin{tabular}[c]{@{}l@{}}Requirement was a workaround for customer requirement and not exactly \\ what customer requested which also resulted in a new requirement instead \\ of the workaround\end{tabular} & \begin{tabular}[c]{@{}l@{}}Deletion + \\ Addition\end{tabular}                     & Remote                                                                                   \\
P18          & Enhancing the product to include a new requested requirement                                                                                                                                                   & Addition                                                                           & Remote                                                                                   \\ \bottomrule
\end{tabular}%
}
\end{table*}

\subsubsection{Intervening Condition: Mode of Work} 
\textit{--Shared by P11--P18.}

Table~\ref{tab:conditions} indicates the modes of work (in--person, and remote) of our participants, which we captured through the pre-interview questionnaire. The majority of our participants worked in--person (N=11) and the rest worked remotely (N=7). There was no hybrid work at the time of data collection.
Our STGT analysis shows that the mode of work intervenes in the execution of the strategies, hence it is an intervening condition. For example, remote work makes it difficult to have open and regular communication, but in--person work makes that easy. For instance,
\begin{quote} \small
    \textit{``I think the face to face has more high bandwidth. But there are lots of circumstances where remote or just not real time is more efficient and more effective. I think, in difficult conversations I prefer face to face. I think it's easier to have a difficult conversation face to face...'' -- P13}
\end{quote}

Similarly, empathising with others is claimed to be more difficult when working remotely, and easier when working in--person. On the other hand, tracking commitments and decisions is claimed to be easier in remote work and more difficult when working i--person. As P17 said,
\begin{quote} \small 
    \textit{``So the communication is pretty transparent. Now that we're doing this remotely. Because it can track any sort of communication that we're doing with the decision making and reasoning. So we always Zoom recordings, so that they are decision driven meetings. Also there are like Slack conversations like there are team channels where there is like a lot of to and fro  between people who are making the important decisions. And so the rest of the team you can also be about, the reasoning why these decisions are being made as well. So in terms of data being transparent, I feel it's much more better when we're doing it remotely, rather than in person when we just have meetings and boards and all. Sometimes there's no note taker, or sometimes we are not actually recording it. So there are no meeting notes to refer back to. So a lot of times important information stages get forgotten after some point in time, but when you do it remotely, you have like an entire history of conversations that happen in the decisions.'' -- P17}
\end{quote}

\subsection{Causes and Consequences}

EI is the ability to process emotional information and use it in reasoning and other cognitive activities. We identified being aware of one's own emotions, and being aware of others' emotions, as the causes of using EI during RC handling; and regulating own emotions, and managing relationships, as the direct consequences of the causes. Additionally, we identified sustaining productivity, and setting and sustaining team goals, as extended consequences. In this section, we describe causes and consequences together, and extended consequences separately.

\subsubsection{Being Aware of Own Emotions (cause) and Regulating Own Emotions (direct consequence)}
\textit{--Shared by all participants.}

Software practitioners feel a wide range of emotions when handling RCs. These span across both high and low pleasurable emotions \cite{Madampe2022TheEngineering}. The common emotions our participants mentioned that they feel during RC handling were \textit{excited, at ease, calm, content, relaxed, anxious, annoyed, frustrated, nervous, stressed,} and \textit{depressed}. Additionally, some participants mentioned the intensity of the emotion they felt, such as very mildly anxious. This indicates that the participants had an awareness of their emotions.

Apart from the emotions they felt, they also mentioned why it is important to be aware of their own emotions when handling RCs. These include:  (a) understanding that emotions are 
influenceable/inheritable, (b) 
understanding that it is not necessary to expresses negative emotions always, and (c) not allowing own negative emotions to impact team goals; which are collectively, \textit{regulating own emotions}. For example,
\begin{quote} \small
    \textit{So when the team is anxious, I also get a little bit anxious, I need to hide that and sort of address the issue with the excitement and energy that I have. Because if I address a team being anxious, the team will sort of inherit that and they will start being very reactive. And they might even not show up for a daily Scrum if they get like that -- P12}
\end{quote}

Our participants stated that emotions influence their communication with others through the tone, especially in the case of seniors in the team, as they influence others via communication such as talking. Also, their actions have an outsized impact on subordinates in the team; therefore, it is important to be aware of own emotions to mitigate inheritance of negative emotions. 
\begin{quote} \small
 \textit{I've seen the value that having someone who is particularly someone more senior in the project who is aware of their feelings, has been able to guide the rest of the team, kind of through that experience of the requirements changes. -- P15}
\end{quote}

\subsubsection{Being Aware of Others' Emotions (cause) and Managing Relationships (direct consequence)}
\textit{--Shared by all participants.}

All of the participants in our study worked in team contexts. In team contexts, there exists a high chance for individuals to witness others' expressing their emotions. Our participants were able to identify the emotions that others in their teams felt when handling the RCs. According to our findings, the main emotion the participants identified in others was \textit{anxiety}. They were also able to recognise differences in emotions others felt according to their role and the level of seniority. As P15 shared,
\begin{quote} \small
    \textit{``there was a very senior member in the team who was just like very calm and very clear and kind of helped the team navigate that emotional journey'' -- P15}
\end{quote}

The key reasons why it is necessary to be aware of others' emotions in handling RCs, as mentioned by our participants, were (a) understanding the influence of others' emotions on decision making, (b) fostering better communication, and (c) to empathise better; which collectively helps in \textit{managing relationships}. 
Our study participants believed that not understanding others' emotions properly can lead to wrong decisions. Thus it is necessary to gauge team members' feelings to understand and question their decisions, and to understand  any individual biases in making decisions.
They also mentioned that for better communication it is necessary to know if others are emotion-driven or not, so that they can potentially tailor their approach to working with them.
As per our analysis, being aware of others' emotions helps empathise better. That is, it helps them to identify what triggers different emotions of others and to interact with others in the team with an understanding. It also helps them to better understand how `the team is feeling' and helps them to better perceive the work the team is doing. They can also get an idea if others understand what they are doing, and see if the team shares the same feelings to maintain the project velocity. For example, as P17 mentioned, 
 
\begin{quote} \small
    \textit{``That is something that I've recently learned that it is important to know how others are  feeling as well. Because at the end of the day, the velocity of the work that we're doing as a team, not just depends upon you, but it depends upon the rest of the team as well, so unless everybody is like pretty enthusiastic about the work we're doing, things do not get done. So it's really important to be aware of how the rest of the team is feeling as well.'--P17}
\end{quote}

\subsubsection{Extended Consequences}

We found the direct consequences of regulating one's own emotions and managing relationships helps in sustaining productivity and setting and sustaining team goals.

\textbf{Sustaining Productivity.} 
\textit{--Shared by P1, P6, P8, P11--P18.}
Our participants said that not being able to regulate their negative emotions impacts their productivity during RC handling. For example, spiraling of emotions (feeding the negative emotions back to themselves to make the situation worse) and the team feeling overwhelmed can harm productivity. They also mentioned that if they felt that they lacked autonomy, did not feel a sense of project ownership, and/or were not able to change this situation, this resulted in lower productivity levels. However, according to our findings, being able to regulate their emotions does not cause a negative impact on their productivity. For example, if they focus on learning from the RC  what they could do better next time (S6), this helps prevent their negative emotions from impacting their productivity. They also stated that emotions can lead to both improved or reduced productivity, depending on how they are handled. Thus they consider emotions as a catalyst and use them to improve productivity. As P15 said,
\begin{quote} \small
\textit{``Sometimes those emotions are good catalysts. I think the the nervousness, as I said, kind of catalysing me to take this discussion and kind of have I've had this revisit of what we've decided previously, like, I think that was like productive because without the emotion without that sort of need to protect the team, I otherwise would have just stuck to the plan and kind of had, you know, inertia carry me forward. So yeah, I think it's interesting. I think they can be both productive and unproductive, depending on how their channels handled.'' -- P15}
\end{quote}

\textbf{Setting and Sustaining Team Goals.} 
\textit{--Shared by P1, P2, P13, P15, P17.} 
Having better relationships and regulating negative emotions were claimed to help set and sustain team goals. For example, as P13 mentioned, at the beginning of the project, the team gets together and decides the direction of the project, i.e., sets a team goal which they do via open communication. And as P2 said,
\begin{quote} \small
    \textit{``We try to keep emotion out of what we do because emotion is not helpful in situations where you have an end goal to deliver'' -- P2}
\end{quote}

\subsection{Contingencies: Strategies}

We identified twelve strategies (given in Table \ref{tab:covariances}) that our software practitioners use to 
support EI during RC handling. Namely, sharing  how individuals feel with manager/ peers (one--to--one/ many) (S1), open and regular communication (many--to--many) (S2), building personal rapport with others (S3), empathise (S4), being open to RCs (S5), focusing on learning from the RC (S6), pre/ post control of expectations about the RC (S7), understanding/ using the level of autonomy an individual has over RC implementation (S8), tracking commitments and decisions (S9), having a dedicated lead for RC (S10), having team/ social rituals (S11), and working as a single cross functional team (S12). These strategies help in being aware of their own emotions (S1, S2), being aware of others' emotions (S2--S4, S11, S12), regulating own emotions (S1, S2, S5--S8),  and managing relationships (S2, S3, S9--S12). 

\subsubsection{Sharing How Individuals Feel with Manager/ Peers (one--to--one/ many) (S1)}
\textit{--Shared by P4, P5, P8, P9, P11--P18.}

This strategy not only helps in being aware of one's own emotions, but also helps in regulating one's own emotions. As our participants mentioned, they share how they feel with others individually and in their team when necessary by talking, messaging, having regular catch ups with the manager and at weekly, monthly, and quarterly frequencies. For example, 
\begin{quote} \small
   \textit{ ``And then like if that helps you realise and understand yourself a bit more I think having someone you can vent to, or write a message to or talk to about like a difficult situation is, is one of the most useful things, ways I've dealt with it.'' -- P13}
\end{quote}

However, some participants mentioned that it is difficult for them to share their emotions openly in group settings, especially during agile retrospectives. Some also mentioned that junior members and new members find it difficult to express their emotions in group settings, where they second guess if their emotions are valid. On the other hand, participants mentioned that senior members express confidently how they feel. This shows that sharing emotions openly happens when adapted to the team context or the company culture or confidence that comes with the position (senior) and experience. Some participants also mentioned that in large groups or when their degree of psychological safety is low, team members are unwilling to express their ``raw'' emotions in front of others in the team. In contrast, they said that small groups better promote the freedom to express these emotions. Therefore, sharing how one feels in one-to-many situations is subjective. However, some participants mentioned that they anonymously share how they feel during agile retrospectives, and retrospectives help reframe emotions to get them out of doom spirals.

\subsubsection{Open and Regular Communication (many--to--many) (S2)}
\textit{--Shared by P1, P2, P4--P18.}

Open and regular communication strategies serve in many ways. It helps being aware of own emotions, being aware of others' emotions, regulating own emotions, and managing relationships. This strategy is very much connected to strategy S1. Our participants mentioned that they have quick discussions during a sprint meeting on resource allocation for the RC, discuss the complexity of existing tasks to decide on RC assignment, discuss team velocity, communicate the mistakes with others in the team, identify risks open to the team very clearly, and question the rationale for RC/ last minute RC through open and regular communication. They mentioned that it helps them ``voice out'' their opinions and feelings, and therefore helps in being aware of own emotions, and regulate their emotions. They also mentioned that open communication with a reduced number of communication layers promotes better communication about RCs, and clarity about the RC. For example,
\begin{quote} \small
    \textit{``Perhaps the problems come in when you have more layers between managers and software developers and the customer and those layers don't have as much an understanding like it's going through a marketing department. I did have this happen to me in previous role marketing department accept changes without shipping the timeline and then feed that down. And that's when you get anger and frustration because that's when you start going to like, we're not working together anymore. Perhaps that's the key phrases. when you're all working together change doesn't matter.'' -- P16}
\end{quote}
 Another such occasion is having weekly one-on-ones with another team's architect where multiple teams work together on an RC. Teams also tend to ask questions from the technical lead during the project, and the technical leads/ managers provide input where necessary to the team. The participants mentioned that they use voice conversations for complex issues and for fast feedback. Additionally, they have quick calls on Slack to discuss if the messaging threads are long. They also mentioned that they openly do Q\&A sessions to discuss decisions made by management when the team had less autonomy in decision making, and they discuss timeline impacts and timeline shifts. This further enhances relationships among members in the team and with the manager. Open and regular communications also allow members to express how they feel freely, making others in the team aware of how they feels. 

\subsubsection{Building Personal Rapport with Others (S3)}
\textit{--Shared by P2, P4--P6, P12, P14, P18.}

Building personal rapport with others helps being aware of others' emotions during RC handling and also better manage relationships. Our participants mentioned that they build personal rapport through talking to identify emotionally driven people, and also they would like to build relationships proactively ahead of the project. As P13 mentioned,
\begin{quote} \small
\textit{``..proactively building relationships ahead of a project. So knowing that a project is coming up in one or two quarters, starting to build relationships with people now so that when I do have to come to them and say, Hi, I need help with this project. They already know who I am. They already know the kind of stuff that I'm going to need from them.'' -- P13}
\end{quote}
Some managers also mentioned that they build personal rapport only where necessary. One non-manager mentioned that they build personal rapport with necessary people only to a limited degree.

\subsubsection{Empathise (S4)}
\textit{--Shared by P3--P5, P8, P10--P16.}

Empathising leads to being aware of others' emotions. Empathising during RC handling divides into three sub categories: empathise (emotion); empathise (cognition); and empathise $\rightarrow$ take action $\rightarrow$ help resolve. According to our findings, all three can be seen in managers, and team members have the first two. 

\textit{Empathise (emotion):} This is where  managers empathise with their team members. Managers try to understand/perceive team's emotions via team leads by asking about the vibe of the team, team members having emotional empathy towards peers, and towards the feature leader. For example, one way that managers  empathise is observing the emotion dynamics (change in emotions over time) of their subordinates. They do this via weekly one on one meetings where they dig deep into team members' emotions. As shared by P8,
\begin{quote} \small
    \textit{``people who are experienced, actually worked, coaching, working one-on-one and just being empathetic to the people in the team.'' -- P8}
\end{quote}

\textit{Empathise (cognition):} This is understanding the rationale for an emotion. For example, understanding why the team members feel negative emotions, understanding/perceiving why a feature lead (dedicated lead of the RC) is getting anxious due to release date shifts, a team's disappointment on priority shift due to RC, understanding/perceiving some members in the team having negative emotions opposite to their own, understanding/perceiving when members who are not familiar with the RC are a little fatigued, and especially managers understanding team members' positions in life and work. P4 mentioned,
\begin{quote} \small
    \textit{``You know, I'm quite open with my, like last week I was just filthy grumpy and I openly said to my team, I said look, I'm just seriously grumpy. It's when I was starting to get sick, and I was like I'm seriously grumpy, I'm actually just going to go home.  And they were like rest, enjoy, just have the afternoon off, you know, because they knew that was what was best for me'' -- P4}
\end{quote}

\textit{Empathise $\rightarrow$ take action $\rightarrow$ help resolve:} According to our findings, this is seen in managers. Managers help others understand emotions/avoid channelling emotions to decisions, take necessary actions to minimise their negative emotion arousal, and listen and validate why they have concerns. For example, managers console team members who are feeling down, help others to move towards the goal rather than thinking about negative emotions such as frustration, they talk with team members about their emotions and how to navigate them, weigh cost over benefits to determine the decision to project the team's feelings, let subordinates express their concerns without them feeling the need to justify the concern, and put in contact with people who can help subordinates with solving the issue they have, or, if possible, step in to resolve the issue with regard to the RC. For example,
\begin{quote} \small
    \textit{''trying to learn people's patterns and drivers so that we can give those opportunities as early, and you know, this particular engineer, nine times out of 10, if we just leave him alone for a couple of hours he solves his problem. And so that's his modus operandi, he's like argh, and then he just goes away, thinks, and then he solves his problem. And we're like that's awesome, because we know that and we can support him, you know, hashtag united, and say here's what we can do to support.'' -- P4}
\end{quote}

\subsubsection{Being Open to RCs (S5)}
\textit{--Shared by P1, P2, P4--P10, P13, P16, P18.}

According to our findings, one of the key strategies for managing one's own emotions is being more open to RCs. Understanding RCs as a natural phenomenon in agile software development contexts and considering them a normal practice help the practitioners be open to RCs, and thereby better manage any negative emotions that could arise due to new RCs. 

\begin{quote} \small
\textit{``When we look when we discover edge cases, like I feel like it's, as I said, it's sort of normal practice for us, like we expect to discover these things. So there's nothing really jarring about it'' -- P14}
\end{quote}

Our participants also mentioned that RCs could be discovered during development, due to technical reasons, edge cases, or even when assumptions are wrong. However, understanding that there is always a reason for the emergence of an RC, and that handling new RCs is for the best of the product, also helps being open to RCs. 

\begin{quote} \small
	\textit{''knowing that there is a reason even if you don't know what it is, there is a reason. I guess it is sort of assuming best intent that you know, a customer wants their software they haven't just done this on a whim.'' -- P16}
\end{quote}

\begin{quote} \small
\textit{``...I don't feel like I have really had negative experiences with requirements changes. And I think that's true to the fact that every time they come through. You understand. Like, it comes with a reason why. And if you understand the reason why I think we pretty much usually it's fine.'' --P13}
\end{quote}

\subsubsection{Focusing on Learning from the RC (S6)}
\textit{--Shared by P1--P5, P9, P13.}

Our participants said that focusing on learnings from their RCs allowed them to manage their emotions during RC handling. For example, RCs could have emerged due to failures/ mistakes. In such occasions, considering the failures/ mistakes as lessons, and understanding the benefits of doing so, have helped our participants to manage their emotions. As P13 said,
\begin{quote} \small
    \textit{``Understand that with every setback, there is a learning and the learning is really what you work with.'' -- P13}
\end{quote}

RCs also could lead to them learning new information, and our participants said that they use the new learnings to steer the decisions, which is an additional benefit. One way of steering the decisions through learnings, as mentioned by P15, is by having quick learnings through experiments and changing decisions quickly. 

\subsubsection{Pre/ Post Control of Expectations about the RC (S7)}
\textit{--Shared by P1, P2, P4, P6, P7, P9, P14, P16, P18.}

According to our findings, practitioners may curate their expectations proactively to avoid low pleasurable emotions. They put in effort to avoid the problem causing low pleasurable emotions by removing the unknowns, and reframe expectations to get past the low pleasurable emotions. For example,
\begin{quote} \small
    \textit{``It's not like oh my god, I've got to change, I've set these expectations. We are very good at treating releases as fluid, you know, when they change it's like we just sort of have this perception, it's not like we change it because we just can't be bothered doing something.'' -- P4}
\end{quote}

\begin{quote} \small
    \textit{``I think when I first started out, it would freak me out, and I'd think, oh what could, how did I not see this, but you can't, you can't plot everything.'' -- P7}
\end{quote}

\subsubsection{Understanding/ Using the Level of Autonomy an Individual has Over RC Implementation (S8)}
\textit{--Shared by P2, P3, P7, P10, P12, P14, P15, P18.}

Understanding/using the level of autonomy an individual has over RC implementation can assist practitioners manage their emotions. Our participants described several aspects related to autonomy, which help them manage their emotions when handling RCs: having autonomy to align with the reason for RC, learning that autonomy can affect change and help deal with RCs, understanding the impossibility of controlling the situation/changing decisions about RC as an individual in certain occasions, and making situations better when there is less autonomy. In addition, the entire team learning the boundaries where they can affect change, i.e., understanding the level of autonomy they have, also aids in managing emotions during RC handling. For example,
\begin{quote} \small
    \textit{``I think autonomy like this part of autonomy is actually important. Because when you understand why, and you're actually aligned with the reason why you're more likely to to not care about the fact that this thing has changed. You're just going to try and solve it.'' -- P13}
\end{quote}

\subsubsection{Tracking Commitments and Decisions (S9)}
\textit{--Shared by P4, P8, P10, P14, P17.}

Tracking commitments and decisions throughout the process of RC handling aids in managing team relationships. This is also supported by the use of appropriate tools. For example, our participants mentioned that they use Jira tickets to track commitments, and document/record decisions using Confluence  when asynchronous communication within the team exists, hence helps in managing the relationships within the team. As P10 said,
\begin{quote} \small
    \textit{``We use confluence for a lot of our changes for that space, depending on the size of the change if it's sort of budget effecting, that's what it really, it's documented quite heavily. But for like changes in my team, it's more the change is just documented as a comment in here'' -- P10}
\end{quote}

\subsubsection{Having a Dedicated Lead for RC (S10)}
\textit{--Shared by P13, P16--P18.}

Having a dedicated lead for a new RC helps to achieve better relationships within the team. According to our findings, having a dedicated lead who is accountable, responsible, and trustworthy helps them have better relationships within the team, as they are the one who sets proper communication within the team and also helps to successfully implement and deliver the RC. For example, as P13 said,
\begin{quote} \small
\textit{``we have what we call feature lead within the team. And essentially, is someone who prays down the stories, runs their solution, runs this mission sessions and make sure that the feature is on track in terms of timeline, and all that kind of stuff. So the visually does all that. I didn't do any of that. So I rely on the feature lead to tell me when they needed me, okay to provide input.'' -- P13}
\end{quote}

\subsubsection{Having Team/ Social Rituals (S11)}
\textit{--Shared by P2, P4, P6, P16.}

Having team/ social rituals is another strategy used by our participants to be aware of others' emotions and maintain better relationships among team members. Some of the rituals our participants mentioned that they have include: having a project pre-mortem to discuss what to expect; having one-on-one coffee sessions within the team; having daily sessions to go through code within the team; and maintaining communication `touchstones' within the team. This strategy allows the team to bond, hence know how others feel, and thereby have better communication and relationships. P16's experience with project pre-mortems as they shared,
\begin{quote} \small
    \textit{``We also as a team like whole series, we have a series of rituals. One quick one that a colleague has brought on is things like Project pre-mortem. Okay. Yeah. So instead of a post-mortem, which, like you've finished your project, back over a pre-mortem is when you try and do that at the start of a project. And you're good to go. Okay, what to go wonderfully, what could go terribly. And everyone's involved in that. So everyone's got a really good idea of what the lay of the land is. So when a change comes, it's never really a big surprise, or it's, everyone was part of the process. So no one feels like, Oh, you shouldn't have got this because there's plenty of opportunity for everyone to have contributed.'' -- P16}
\end{quote}

\subsubsection{Working as a Single Cross--Functional Team (S12)}
\textit{--Shared by P1, P2, P4, P6--P12, P14, P16.}

Working as a single cross--functional team helps team members to be aware of others' emotions and manage relationships. Working on an RC is not a single function task. Various teams may be involved from receiving the RC, to developing and testing, to delivering the RC. One important fact our participants shared with us is that it is necessary to work as a single cross-functional team, especially when it comes to large organisations. 
For example, managers of the teams involved in the RC communicate with each other and receive/ take the responsibility ON the RC, build the sense of community across the people who are working on the product together, and treat teams of all disciplines as a single cross--functional team, so that other teams are not considered as external contributors. This allows one to know how others in a cross-disciplinary team feel, and also to have better understanding and relationships with them.
\begin{quote} \small
    \textit{``What we've done is rather than treating the separate disciplines as external contributors, we treat all of the disciplines as cross--functional single team, okay. And, like I said, build that sense of community across the people who are working on the product together. And in doing that we control for the feeling of the other. And then their change. There's more of an understanding that there's a human on the other side of the change, who was also trying to do the best thing they're trying to do. So vice versa, the engineers have more of an understanding of who the product of the design team are, and why they might be doing the things they're doing. And similarly, the design of the product team, have more of an empathy for the engineers who are saying we can't do that it's not possible and the perfect design that you put together that would look really great.'' -- P13}
\end{quote}

\subsection{Covariances}

Covariance occurs when one variable changes with the changes of the other. In our case, the variables are the six Cs. For example, for different causes (cause is a variable), there are different contingencies: strategies (contingency: strategy is the other variable which changes with the cause). We found three covariances in our study, (A) strategies vary with the causes, (B) strategies vary with the direct consequences, and (C) Ease/ difficulty in executing strategies vary with intervening conditions. We have summarised these covariances in Table \ref{tab:covariances}.

\textit{Strategies vary with causes (Covariance A). } 
The strategies used to be aware of own emotions and to be aware of others' emotions co-vary. However, they also share a common strategy (S2).

\textit{Strategies vary with direct consequences (Covariance B). } 
Similar to covariance A, the strategies used to regulate one's own emotions and to manage relationships co-vary. Common strategies (S2, S4, S9--S12) exist for both direct consequences. 

\textit{Ease/ difficulty in executing strategies vary with intervening conditions (Covariance C). }
Unlike covariance A and B, covariance C consists of a dimension. I.e., the ease/ difficulty in executing the strategies with intervening conditions, which is the mode of work. This covariance applies to the strategies S2, S4, and S9.

\begin{table*}[]
\caption{Covariances}
\label{tab:covariances}
\resizebox{\textwidth}{!}{%
\begin{tabular}{@{}llllllll@{}}
\toprule
\textbf{}          & \textbf{Covariance}                                                                                                                  & \multicolumn{2}{l}{\textbf{A}}                                                                  & \multicolumn{2}{l}{\textbf{B}}                                                                                                         & \multicolumn{2}{l}{\textbf{C}}                                   \\ \midrule
                   &                                                                                                                                      & \multicolumn{2}{l}{\textbf{Causes}}                                                                      & \multicolumn{2}{l}{\textbf{Direct consequences}}                                                                                                & \multicolumn{2}{l}{\textbf{Mode of work}}                                 \\ \cmidrule(l){3-8} 
\multirow{-2}{*}{} & \multirow{-2}{*}{\textbf{Strategies}}                                                                                                & \begin{tabular}[c]{@{}l@{}}Being aware of \\ own emotions\end{tabular} & \begin{tabular}[c]{@{}l@{}}Being aware of \\ others' emotions\end{tabular} & \begin{tabular}[c]{@{}l@{}}Regulating own \\ emotions\end{tabular} & \begin{tabular}[c]{@{}l@{}}Managing \\ relationships\end{tabular} & \begin{tabular}[c]{@{}l@{}}Remote \\ working\end{tabular}               & \begin{tabular}[c]{@{}l@{}}In--person \\ working\end{tabular}                 \\ \cmidrule(r){1-2}
\rowcolor[HTML]{EFEFEF} S1                 & \cellcolor[HTML]{EFEFEF}\begin{tabular}[c]{@{}l@{}}Sharing how individuals feel with \\ manager/peers (one-to-one/many)\end{tabular} & 
\cellcolor[HTML]{EFEFEF}\faCheckSquare                                                     &
\cellcolor[HTML]{EFEFEF}                                                   & \cellcolor[HTML]{EFEFEF}\faCheckSquare                                        & \cellcolor[HTML]{EFEFEF}                                          & \cellcolor[HTML]{EFEFEF}     & \cellcolor[HTML]{EFEFEF}          \\
S2                 & \begin{tabular}[c]{@{}l@{}}Open and regular communication \\ (many-to-many)\end{tabular}                  &               \faCheckSquare            & \faCheckSquare                                                                        & \faCheckSquare                                                                & \faCheckSquare                                                               & Difficult                    & Easy                              \\
\rowcolor[HTML]{EFEFEF} S3                 & \cellcolor[HTML]{EFEFEF}\begin{tabular}[c]{@{}l@{}}Building personal rapports with \\ others\end{tabular}        &          \cellcolor[HTML]{EFEFEF}           & \cellcolor[HTML]{EFEFEF}\faCheckSquare                                                & \cellcolor[HTML]{EFEFEF}                                           & \cellcolor[HTML]{EFEFEF}\faCheckSquare                                       & \cellcolor[HTML]{EFEFEF}     & \cellcolor[HTML]{EFEFEF}          \\
S4                 & Empathise                                                                                                         &                   & \faCheckSquare                                                                        &                                                                    &                                                                   & Difficult                    & Easy                              \\
\rowcolor[HTML]{EFEFEF} S5                 & \cellcolor[HTML]{EFEFEF}Being open to RCs                                                      &                         \cellcolor[HTML]{EFEFEF}             &  \cellcolor[HTML]{EFEFEF}                                                   & \cellcolor[HTML]{EFEFEF}\faCheckSquare                                        & \cellcolor[HTML]{EFEFEF}                                          & \cellcolor[HTML]{EFEFEF}     & \cellcolor[HTML]{EFEFEF}          \\
S6                 & Focusing on learning from the RC                                                                                                     &                                                    &                        & \faCheckSquare                                                                &                                                                   &                              &                                   \\
\rowcolor[HTML]{EFEFEF} S7                 & \cellcolor[HTML]{EFEFEF}\begin{tabular}[c]{@{}l@{}}Pre/post control of expectations \\ about RC\end{tabular}             &        \cellcolor[HTML]{EFEFEF}    & \cellcolor[HTML]{EFEFEF}                                                   & \cellcolor[HTML]{EFEFEF}\faCheckSquare                                        & \cellcolor[HTML]{EFEFEF}                                          & \cellcolor[HTML]{EFEFEF}     & \cellcolor[HTML]{EFEFEF}          \\
S8                 & \begin{tabular}[c]{@{}l@{}}Understanding/using the level of \\ autonomy individual has over RC \\ implementation\end{tabular}     &   &                                                                            & \faCheckSquare                                                                &                                                                   &                              &                                   \\
\rowcolor[HTML]{EFEFEF} S9                 & \cellcolor[HTML]{EFEFEF}Tracking commitments and decisions                                                             &      \cellcolor[HTML]{EFEFEF}        & \cellcolor[HTML]{EFEFEF}                                                   & \cellcolor[HTML]{EFEFEF}                                           & \cellcolor[HTML]{EFEFEF}\faCheckSquare                                       & \cellcolor[HTML]{EFEFEF}Easy & \cellcolor[HTML]{EFEFEF}Difficult \\
S10                & Having a dedicated lead for RC       &                                                                                                &                                                                            &                                                                    & \faCheckSquare                                                               &                              &                                   \\
\rowcolor[HTML]{EFEFEF} S11                & \cellcolor[HTML]{EFEFEF}Having team/social rituals                                            &                  \cellcolor[HTML]{EFEFEF}                     &  \cellcolor[HTML]{EFEFEF}\faCheckSquare                                                & \cellcolor[HTML]{EFEFEF}                                           & \cellcolor[HTML]{EFEFEF}\faCheckSquare                                       & \cellcolor[HTML]{EFEFEF}     & \cellcolor[HTML]{EFEFEF}          \\
S12                & \begin{tabular}[c]{@{}l@{}}Working as a single cross-functional \\ team\end{tabular}                &                                 & \faCheckSquare                                                                        &                                                                    & \faCheckSquare                                                               &                              &                                   \\ \bottomrule
\end{tabular}%
}
\end{table*}

\section{Discussion}
In this section, we discuss the implications of our findings for practitioners, including recommendations, and implications for researchers with corresponding ideas for future work.

\subsection{Implications for Practitioners}

\textbf{Recommendation 1. Communicate openly and regularly.}
As shown in Table \ref{tab:covariances}, open and regular communication is the only strategy that can assist in achieving all four aspects of EI. Therefore, it is highly beneficial if software practitioners practice open and regular communication during RC handling.

\textbf{Recommendation 2. Embrace RCs.}
RCs are unavoidable in software contexts. According to our findings, resisting RCs does not help, but in contrast, embracing them helps in improving EI when handling RCs. Consequently, this helps in improving the product delivery as well.

\textbf{Recommendation 3. Be team players.}
Half of the strategies that we found (S1--S4, S11, S12) can be said to be inter--personal strategies. Therefore, improving EI is necessary not only from an individual perspective, but also to be a better team player. 

\textbf{Recommendation 4. Junior members and new members: don't be afraid to share what you feel with others when handling RCs.} 
Our findings show that junior members and new members tend not to share how they feel with others, and second guess the emotions they feel. While sharing emotions in group settings depends on whether the person feels comfortable, sharing how one feels with others helps in improving EI when handling RCs. This also highly relies on the psychological safety within the team.

\textbf{Recommendation 5. Use appropriate tools and their features to execute strategies according to the mode of work.}
Our findings indicate that execution of strategies is impacted by the mode of work. Software teams can think how to avoid the mode of work (remote/ in--person) being an intervening condition. For example, P13 mentioned that they use the ``huddle'' feature in Slack to have quick calls instead of instant messaging when the discussions are long. Such features can be used in remote working settings to enhance open and regular communication.

\textbf{Recommendation 6. Use the problem--solution chart we have presented: Strategies to use during RC handling to support emotional intelligence.}
In Table \ref{tab:covariances}, the strategies we have listed can be used during RC handling to help team members raise awareness of their own emotions, to be aware of others' emotions, to regulate their own emotions, and to help manage their relationships. This table can be used as a problem--solution guide. For example, to be aware of others' emotions, one can consider using the strategies, sharing how individuals feel with manager/ peers (S1), or open and regular communication (S2).

\subsection{Implications for Researchers}

\textbf{The theoretical framework.}
Fig. \ref{fig:sixCs} outlines our theoretical framework for improving EI during RC handling. This framework can be used to get a better understanding about the central phenomenon of emotional intelligence of software practitioners during RC handling.

\textbf{ What about the strategies used by practitioners on the other side of the world?}
The eighteen interviews we conducted saturated what we found, and the findings are limited predominantly to the Australasian region, possibly exhibiting territorial, organisational and cultural specificities. Researchers may replicate our study in other regions of the world to see if the practitioners use the same strategies, or whether they use different strategies. Researchers can then expand our theoretical framework to bring more understanding and knowledge on the central phenomenon to include other parts of the world as well.

\textbf{Applicability of strategies in situations beyond RC handling.}
The strategies we found are limited to RC handling. These could be applicable for situations beyond RC handling. Researchers may consider studying this further by conducting a study such as a survey to know if the strategies we have listed can be used in other scenarios as well.

\textbf{Mode of work intervening in the executing of all strategies.}
We only have found how mode of work intervenes executing some of the strategies we found, i.e., how the mode of work intervenes S2, S4, and S9. A future study can consider if the mode of work intervenes the execution of other strategies as well.

\section{Evaluation and Limitations} 

\textbf{Evaluation.} We evaluate our method application against credibility and rigour, and our outcome against originality, relevance, and density, as per the STGT evaluation guidelines \cite{Hoda2021Socio-TechnicalEngineeringb}. \textit{Credibility:} We have provided details on how participants were recruited (social media, personal contacts, and from a large software company), the applied sampling method (convenience sampling and theoretical sampling), how iterative and interleaved data collection and analysis occurred, and the memos written (textual: interview reflections, conceptual memos; visual: Miro board). \textit{Rigour:} We have provided an example of open coding, and constant comparison, embedded sanitised evidence (quotes). \textit{Originality:} To the best of our knowledge, this is the first study on EI in handling RCs (confirmed through related work). \textit{Relevance:} Our previous studies (\cite{Madampe2022TheEngineering}, \cite{Madampe2022Emotion-CentricEngineering}) prove how relevant the central phenomenon is for RC handling contexts. \textit{Density:} The findings are layered and full of real examples (quotes).

\textbf{Limitations.}
In our study, we collected data from eighteen software practitioners engaged in requirements change handling in software engineering. While this is a relatively small number of participants, each interview was between 30--60 minutes long, sharing detailed accounts and experiences on the relevant issues. Qualitative analysis of these interviews provided rich insights into the various components of EI as it applies to requirements change handling, as is evident from the multi-faceted nature of the theoretical framework. The participants represented the Australasian region and among them the majority were from Australia and New Zealand. Therefore, the applicability of the findings is limited to the contexts studied in these regions. Future work can study emotional intelligence in other contexts and expand the applicability of our findings. When working in a team, coding by a single researcher can lead to a potentially restricted view of the data. To address this, the first author analysed all data, shared the findings with the second and third authors. During weekly meetings, the second author went through the codes of the first iteration to confirm the reliability. In \circled{2}, the first author wrote extensive memos of interview reflections and conceptual memos, visualised findings using Miro, and shared with both the second and third authors during the process. We also had fortnightly discussions on findings.

\section{Conclusion}
We set out to study the role of emotional intelligence during requirements change handling in software engineering. Through an empirical study, we identified the six Cs of software practitioners' emotional intelligence during requirements change handling, including the \textit{context}, \textit{conditions}, \textit{causes}, \textit{consequences}, and \textit{contingencies} (strategies). We found that there exist \textit{covariances} among causes, consequences, and strategies, and mode of work intervenes the execution of the strategies we identified. These are inclusive of, but also go beyond, the four known aspects of emotional intelligence -- self-awareness (being aware of own emotions), social awareness (being aware of others' emotions), self regulation (regulating own emotions), and relationship management (managing relationships) \cite{Goleman2002TheTeams}. Based on our findings, we provide a problem--solution chart that can be used to choose the most appropriate strategy to be aware of one's own emotions, aware of others' emotions, regulate own emotions, manage relationships, whether the team is working remotely or in--person. Further, we present a set of recommendations for software practitioners, and a set of potential future work for researchers.

\bibliographystyle{ieeetr}
\bibliography{main}

\ifCLASSOPTIONcompsoc
  \section*{Acknowledgments}
  This work is supported by a Monash Faculty of IT scholarship. Grundy is supported by ARC Laureate Fellowship FL190100035. Our sincere gratitude goes to the large software company who gave us the opportunity to recruit participants, and 
  to everyone at the company who helped with reviewing the pre--interview questionnaire, recruiting participants, reading the draft and providing valuable feedback, and to all the participants who took part in this study.
\else
  \section*{Acknowledgment}
\fi

\ifCLASSOPTIONcaptionsoff
  \newpage
\fi



%

%

\appendices

\section{Memos}

\begin{strip}
             \begin{tcolorbox}[width=\textwidth,colback=gray!10, boxrule=0pt,frame hidden]
			\textbf{Interview Reflection (Written right after the interview with P17)}
			
			The participant is a new recruit. They said they got confused when changes happen last minute. Also, relaxed when a requirement which was not exactly what customer requested got removed. 
            They also appreciated   ``transparent communication'', which is the open communication that everyone talked about. Openness seems to be contributing a lot to keep the participants positive during requirements change handling. Open communication seems to belong to both own emotion regulation and managing relationships.
            The participant likes to work remotely - has joined the company remotely as well. They believe that information losses are prevented when working remotely as all decisions, and all communications, are recorded (recording decisions). P13 mentioned documenting all details on RCs as well. Documentation seems to be important in all cases.
            They also mentioned with date shifts, the feature lead is especially anxious. P16 is a feature lead, and they talked a lot about date shifts. But as P16 is an experienced person, according to what they said, they seem to be handling the negative emotions well as they accept the concept of RCs, and use open communication.
            The key ideas at this point by looking at all interviews conducted up to now during iteration 2 are, \textbf{accepting RCs as a natural phenomenon} helps to regulate one's own emotions, \textbf{open communication} helps to regulate one's own emotions and enhances better relationships within the team and beyond, and all teams involved from various disciplines \textbf{working as a single cross-functional team} helps maintain the positivity as an individual/ as an organisation. \\
            
            \textbf{Conceptual Memo (Written about the condition: Mode of work)}
            
            There are pros and cons of remote and in--person RC handling. Empathising is difficult in remote working, but easy when handling the RCs in--person. Empathising is used as a strategy in being aware of others' emotions. Similarly, ``enables recording decisions'' is a pro in remote RC handling and ``losing information in decision making'' is a con in in--person RC handling. But recording decisions is a strategy used to manage relationships. The same applies to open communication. So, the mode of work intervenes in the execution of the strategies.
            \end{tcolorbox}           
\end{strip}

\begin{IEEEbiographynophoto}{Kashumi Madampe}
is a final year PhD candidate at Monash University, Melbourne, Australia. 
Ms. Madampe did part of her current research at The University of Auckland, New Zealand. Prior to the PhD candidature, she was a software project manager and a business analyst. 
Her research interests are software engineering, agile, human--computer/AI interaction, grounded theory, and natural language processing. She serves as a peer reviewer for ACM Computing Surveys Journal, Automated Software Engineering Journal, Journal of Systems and Software, and in several conference program committees. She also served as the XP2021 poster co--chair, ASE2021 and CHASE2021 publicity and social media co--chair. More details about her research can be found at https://kashumim.com. Contact her at kashumi.madampe@monash.edu.
\end{IEEEbiographynophoto}

\begin{IEEEbiographynophoto}{Rashina Hoda}
is an Associate Professor in Software Engineering at the Faculty of Information Technology, Monash University, Melbourne. Her research focuses on human and social aspects of software engineering, socio-technical grounded theory, and serious game design. She serves on the IEEE Transactions on Software Engineering review board, IEEE Software advisory board,
as ICSE2021 social media co--chair, CHASE
2021 program co--chair, and ICSE2023 SEIS co-chair. For details see www.rashina.com. Contact her at
rashina.hoda@monash.edu.
\end{IEEEbiographynophoto}


\begin{IEEEbiographynophoto}{John Grundy}
received the BSc (Hons), MSc, and PhD degrees in computer science from the University of Auckland, New Zealand. He is an Australian Laureate fellow and a professor of software engineering at Monash University, Melbourne, Australia. He is an associate editor of the IEEE Transactions on Software Engineering, the Automated Software Engineering Journal, and IEEE Software. His current interests include domain--specific visual languages, model--driven engineering, large-scale systems engineering, and software engineering education. More details about his research can be found at https://sites.google.com/site/johncgrundy/. Contact him at john.grundy@monash.edu.
\end{IEEEbiographynophoto}




\end{document}